\documentclass[conference]{IEEEtran}
\usepackage{amsbsy}
\usepackage{floatflt} 

\usepackage{amsmath,stackrel}
\usepackage{amssymb}
\usepackage{times}
\usepackage{graphics}
\usepackage{graphicx}
\usepackage{xspace}
\usepackage{paralist} 
\usepackage{setspace} 
\usepackage{xypic}
\xyoption{curve}
\usepackage{latexsym}
\usepackage{theorem}
\usepackage{ifthen}
\usepackage{subfigure}
\usepackage{booktabs}
\usepackage{algorithm}
\usepackage{algorithmic}
\usepackage{color}

\newcounter{brojac}

{\theoremheaderfont{\it} \theorembodyfont{\rmfamily}
\newtheorem{remark}[brojac]{Remark}
}

{\theoremheaderfont{\it} \theorembodyfont{\rmfamily}
\newtheorem{theorem}{Theorem}

}

\ifCLASSINFOpdf
\else
\fi
\begin{document}
%
\title{Slotted Aloha for Networked Base Stations with Spatial and Temporal Diversity}

\author{\IEEEauthorblockN{Du$\check{\mbox{s}}$an Jakoveti\'c, Dragana Bajovi\'c}
\IEEEauthorblockA{BioSense Center,\\
University of Novi Sad,
Novi Sad, Serbia\\
Email: \{djakovet, dbajovic\}@uns.ac.rs}
\and
\IEEEauthorblockN{Dejan Vukobratovi\'c, Vladimir Crnojevi\'c}
\IEEEauthorblockA{Department of Power, Electronics, and Communications Engineering,\\
University of Novi Sad,
Novi Sad, Serbia\\
Email:\{dejanv, crnojevic\}@uns.ac.rs}}


%


\maketitle

\vspace{-0mm}

\begin{abstract}
We consider framed slotted Aloha where $m$ base stations cooperate to decode messages from $n$ users.
Users and base stations are placed uniformly at random over an area. At each frame, each user
sends multiple replicas of its packet according to a prescribed distribution,
and it is heard by all base stations within the communication radius~$r$.
Base stations employ a decoding algorithm that utilizes the successive interference cancellation mechanism,
both in space--across neighboring base stations, and in time--across different slots, locally at each base station. We show that
there exists a threshold on the normalized load $G=n/(\tau m)$, where $\tau$ is the number of slots per frame,
below which decoding probability converges asymptotically (as $n,m,\tau\rightarrow \infty$, $r\rightarrow 0$)
to the maximal possible value--the probability that a user is heard by at least one base station,
and we find a lower bound on the threshold. Further, we give a heuristic evaluation of the decoding probability based
on the and-or-tree analysis. Finally, we show that the peak throughput increases linearly in the number of base stations.

%
%
%
%
%
\end{abstract}



%
\IEEEpeerreviewmaketitle

\section{Introduction}
\label{section-intro}
We study slotted Aloha for systems with multiple base stations, where the user-base station adjacency is induced by the system's geometry.
Users and base stations are deployed uniformly at random over an area.
Each user transmits its packet replicas at multiple slots, according to a prescribed distribution~$\Lambda_q$, and is heard by any base station that lies within distance~$r$ from it.
  We design a decoding algorithm that utilizes the (successive) interference cancellation~(SIC) mechanism
   in two ways--spatially and temporally.
   The spatial SIC mechanism works as follows. Whenever a base station detects a clean signal at a certain slot, it not only decodes the signal and collects a user, but it also informs all other base stations that share (cover) the collected user. Subsequently, each of the receiving stations removes the user's contribution from their local signals which, in turn, potentially yields new clean signals and collects additional users.
    During the temporal SIC, each base station separately runs the standard SIC across different slots, as in, e.g.,~\cite{liva}.
   The algorithm proceeds iteratively by interleaving spatial and temporal SICs.


\begin{figure}[thpb]
      \centering
      \includegraphics[trim = 2.5mm 0mm 0mm 0mm, clip, width=8.8cm]{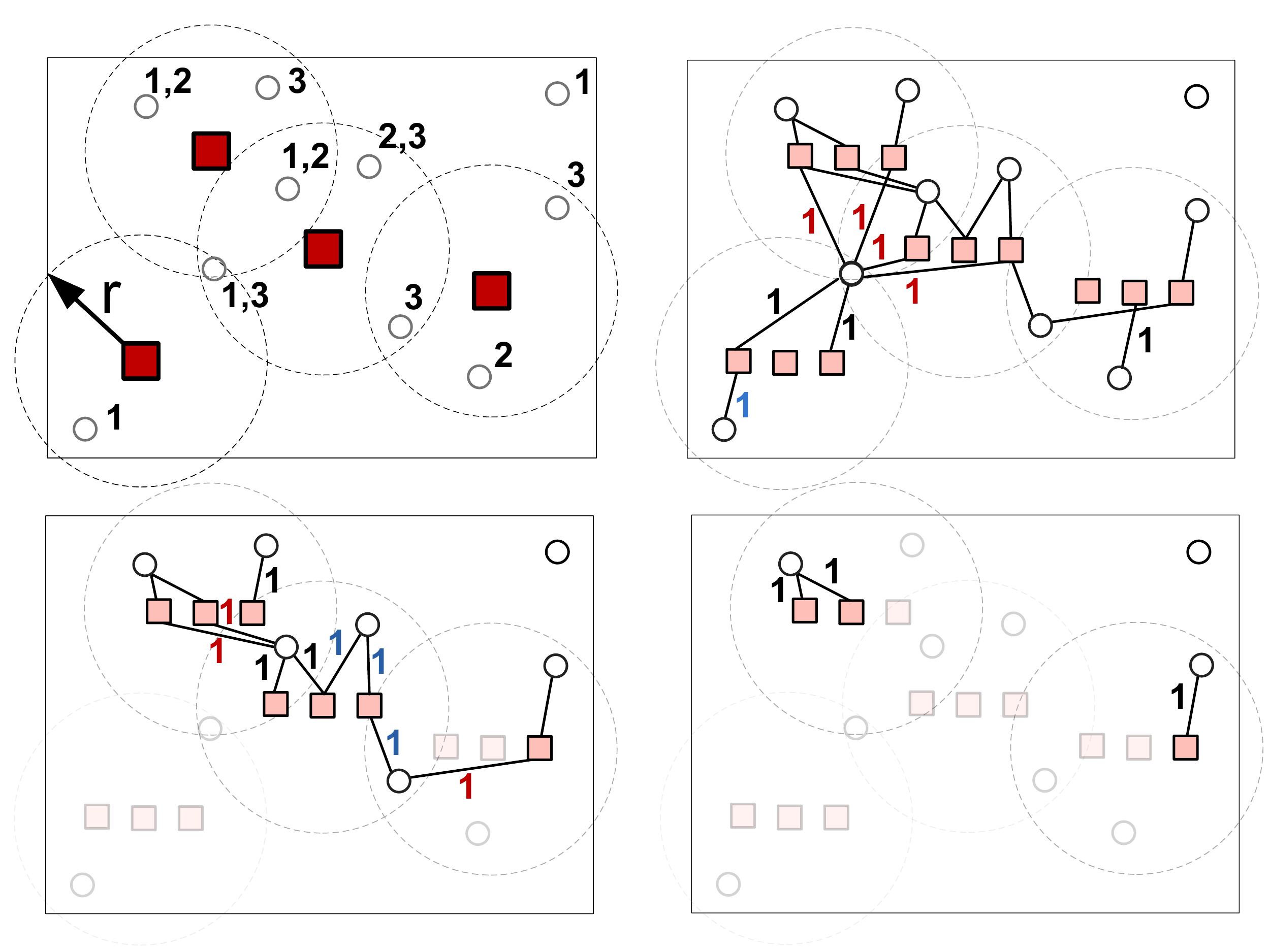}
       \caption{ Illustration of the system with $\tau=3$ slots. The top left figure
       shows the physical system. Red squares represent base stations, and circles
       represent users. Each user is labeled with its activation (transmitting) slots.
       The top right figure depicts the associated graph~$\mathcal G$ defined in Section~\ref{section-model}.
       At each base station, pink squares represent local check nodes for each slot, ordered
       from left to right. The sequence of Figures top right, bottom left, and bottom right, shows
       respectively the iterations $s=1,2,3,$ of the proposed decoding algorithm.
       We label with ``1'' (colored either black, blue, or red) the edges adjacent to
       users decoded at~$s$. Black ``1'' corresponds to the first iteration of a local temporal SIC;
       blue ``1'' corresponds to the second (or larger) iteration of a local temporal SIC;
        and red ``1'' corresponds to spatial SIC.
       }
       \label{figure-example}
       \vspace{-0mm}
\end{figure}

We establish that there exists a threshold~$G^\star$ on the
normalized load $G=n/(\tau m)$ (with $n$ and $m$ respectively the number of users and base stations and $\tau $
 the number of slots) below which the decoding probability converges asymptotically\footnote{Our asymptotic setting
assumes that: $n,m,\tau \rightarrow \infty$; $r\rightarrow 0$; and the users' average degree $\delta$ and the normalized load
$G=n/(\tau m)$ converge to positive constants.} to its maximal possible value, equal to the probability
that a user is heard by at least one base station. Further, we find an explicit lower bound on the threshold~$G^\star$ in terms of: 1) the users’ asymptotic average degree $\delta$; and 2) the
threshold~$H^\star$ on the load (number of users per slot) of the
corresponding single base station system with standard SIC decoding and the same users' degree distribution~$\Lambda_q$.

%
%

%
 %
 Our results reveal that the gains in the peak throughput (average number of collected users per slot, across all base stations) of our $m$-base station system
 are linear in~$m$ with respect to the single base station system. Further, we derive an and-or tree heuristic for decoding probability and show
 by simulation that the heuristic follows well the actual system's performance. Finally, we show by simulations
 that utilizing \emph{simultaneously} spatial and temporal SICs, as proposed here, yields significant gains over decoding schemes
  that employ either only temporal or only spatial SIC (or none of the two). For example, for $\delta=3$, $m=40$, and $\tau=40$, the spatio-temporal SIC achieves peak throughput about $18$, while the other two algorithms achieve at most~$9$.


\textbf{Brief comment on the literature}. Slotted Aloha with single base station and
(temporal) SIC has been proposed in~\cite{SlottedALOHAwithIC} and
further analyzed and optimized in~\cite{liva}. Multiple base stations (receivers) systems have been previously considered in~\cite{ZorziSpatialDiversity}, and very recently in~\cite{LivaNovo}. Both the model from~\cite{LivaNovo} and the one in~\cite{ZorziSpatialDiversity} are very different from ours. Reference~\cite{ZorziSpatialDiversity} studies the capture effect with multiple antennas in the presence of fading and shadowing. Reference~\cite{LivaNovo} assumes independent on-off fading across different user-receiver pairs and derives analytically the decoding probability, when each receiver works in isolation from other receivers.
In~\cite{MASSAP2}, we study the same
geometric setup as studied here, but~\cite{MASSAP2} considers a different decoding, with \emph{spatial SIC only}.
In summary, this paper and~\cite{MASSAP2} differ from existing work by employing cooperation and SIC across
neighboring base stations. With respect to~\cite{MASSAP2}, both the proposed algorithm and the corresponding theoretical contributions are novel.

%

\textbf{Paper organization}. Section~\ref{section-model} describes the system model, and Section~\ref{sec-decoding-algorithm} presents the
proposed decoding algorithm. Section~\ref{section-performance-analysis} presents our results
on the algorithm's performance. Section~\ref{section-numerical} gives numerical simulations. Finally, we conclude in Section~\ref{section-conclusion}.

\section{System model}
\label{section-model}
%
%
%
%
%
We consider a multi-access system with $n$ users and $m$ base stations. (See Figure~\ref{figure-example}, top left, for a system illustration.) We let $U_i$ and $B_l$, respectively, denote user $i$ and base station $l$.  Each user in the system can be served by any of the base stations in its proximity, modelled as a disc of radius~$r$.
  Likewise, the signal that station $B_l$ hears is a superposition of the signals of all transmitting users that are within distance~$r$ from $B_l$. For any $U_i$ and $B_l$ that are at most $r$ apart, we say that $U_i$ and $B_l$ are adjacent.
 Before detailing the system, we introduce here
  our basic terminology and basic notions. Namely, if the signal from $U_i$ is decoded at station~$B_l$, we say that $B_l$ collects $U_i$.
 Also, we say that $U_i$ is collected in the system if it is collected by at least one adjacent base station.

{\bf{Users' and base stations' placements}}. Each user is placed uniformly at random over a unit square $\mathcal A:= \left\{ (x,y) \in \mathbb R^2:\, |x|\leq 1/2,\, |y|\leq 1/2 \right\}$, and independently of other users. We denote the position of $U_i$ by $u_i$. Similarly, each base station is placed uniformly at random over $\mathcal A$, independently of the users' placements, and independently of all other base stations. We denote the position of $B_l$ by $b_l$.
For the purpose of analysis, we differ two types of placements: 1) a nominal placement, when a user or a base station belongs to the interior $\mathcal A^{\mathrm o,r}:=\left\{ (x,y) \in \mathbb R^2:\, |x|\leq 1/2-2r,\, |y|\leq 1/2-2r\right\}$ of $\mathcal A$; and 2) a boundary placement, when a user or a base station belongs to the strip $\partial \mathcal A^{r}:= \mathcal A\setminus \mathcal A^{\mathrm o,r}$ along the boundary of $\mathcal A$.

Let $D_i$ denote the number of base stations adjacent to $U_i$; we call $D_i$ the \emph{spatial degree} of $U_i$.
For any nominal placement of $U_i$, $D_i$ is a binomial random variable with the number of trials $m$ and success probability $r^2\pi$, that is, $\mathbb P\left(D_i= d \, | \, u_i\in \mathcal A^{\mathrm{o},r} \right)= {m\choose d} (r^2\pi)^d (1-r^2\pi)^{m-d}$, for any $d=1,...,m$.
We will be interested in the asymptotic regime $m,n\rightarrow +\infty$, $r\rightarrow 0$,
  with $r=r(n)$ such that $m r^2 \pi =\delta$, for all $n$, and $\delta>0$ is a constant. In this case, conditioned on the nominal placement, the distribution of $D_i$ becomes Poisson with mean~$\delta$:
 $
\mathbb P\left(D_i= d \, | \, u_i\in \mathcal A^{\mathrm{o},r} \right) \rightarrow \Delta_d:= \frac{\delta^d} {d!}e^{-\delta}.
 $
Denote by $\mathbb P \left( U_i\,\mathrm{cov.} \right)$ the probability
that user $U_i$ has at least one adjacent base station, i.e., $U_i$ is covered. Asymptotically,
$\mathbb P \left( U_i\,\mathrm{cov.} \right) \rightarrow 1-\Delta_0=1-\mathrm{exp}(-\delta)$.
Clearly, $U_i$ cannot be collected by any decoding algorithm unless it is covered, i.e.,
$\mathbb P \left( U_i\,\mathrm{coll.} \right) \leq \mathbb P \left( U_i\,\mathrm{cov.} \right)$.

{\bf{Transmission protocol}}. We consider a framed system with $\tau$ slots per frame.
Each user $U_i$ transmits its packet replicas in one or more randomly selected slots.
 We say that $U_i$ is active at a certain slot if it transmits at that slot.
 At each activation slot in a fixed frame, $U_i$'s message contains the information packet (same across all activation slots in the frame),
 $U_i$'s ID, and the list of all activation slots in the frame.
  Users' and base stations' placements are assumed fixed during each frame. Further, slots and frames are synchronized across all base stations, and duration of each slot is the same at all base stations. We thus have, at each frame, global slots $t=1,...,\tau$ common to all base stations.
 Also, users' transmissions are synchronized with respect to slots $t=1,...,\tau$.
 Base stations perform decoding (detailed in Section~\ref{sec-decoding-algorithm}) at the end of each frame (after the users finish the transmission phase in the frame).
   From now on, without loss of generality, we focus on one arbitrarily chosen frame.

We define the normalized load $G$ as the number of users per base station per slot:
  $G:=n/(\tau m)$. Let $Q_i$ denote the number of slots in which $U_i$ transmits in the frame; we call $Q_i$ the $U_i$'s \emph{temporal degree}. The random variable $Q_i$ takes its realizations according to a certain prescribed distribution $\{\Lambda_q, q=1,2,...,\infty\}$:
$
\mathbb P\left(Q_i= q\right) = \Lambda_q,\;\mathrm{for}\,q=1,2,...,\infty;
$
 we accordingly call this distribution the users' \emph {temporal degree distribution}\footnote{In the most general case,
the values $\Lambda_q$ may depend on $\tau$, but with frequently used distributions, like in~\cite{SlottedALOHAwithIC,liva}, they do not depend on $\tau$. As is natural, we assume throughout that distribution~$\Lambda_q$ has a finite support:
 $\Lambda_q=0$ for all $q \geq q_{\mathrm{max}}$, for all~$\tau$, where $0<q_{\mathrm{max}}< \infty$ is a constant independent of~$\tau$.}.
Clearly, $\Lambda_q=0$ for $q>\tau$. 
Also, denote by $\lambda :=\mathbb E [Q_i]$ the expected temporal degree. We finally remark that, once it is decided on a certain realization $q$ of $Q_i$, the $q$ slots in which $U_i$ transmits are chosen uniformly at random.


{\bf{Graph representation of the system}}.
We associate to our system a bipartite graph $\mathcal G=(\mathcal U, \mathcal V, \mathcal E)$.
Graph $\mathcal G$ encodes, for each station~$B_l$ and each slot~$t$, which users
participate in the superposition signal at $B_l$ and slot~$t$.
The variable nodes $\mathcal U = \{U_1,...,U_n\}$ are the users,
and the check nodes are base station-slot pairs $(B_l,t)$, $l=1,...,m$, $t=1,...,\tau$. There are
$\tau m$ check nodes. %
%
 Finally, the set of edges $\mathcal E$ is the set of all pairs $\left(U_i, (B_l,t)\right)$ such that $U_i$ and $B_l$ are adjacent and $U_i$ is active at slot~$t$. See Figure~\ref{figure-example} (top left) for an illustration of the system, and
 Figure~\ref{figure-example} (top right) for the corresponding graph~$\mathcal G$.

{\bf{Degree distributions in~$\mathcal G$}}. Denote by $Z_i$ the degree of $U_i$ (arbitrary variable node) in~$\mathcal G$.
Since the users' and base stations' placements are fixed during the frame, whenever active, $U_i$ is heard by the same set of base stations. Therefore, $Z_i=D_i Q_i$, i.e., $Z_i$
 is the product of the $U_i$'s spatial and temporal degrees.
  Polynomial representation of $Z_i$ is defined by:
$
\mathbb E\left[ x^{Z_i} \right] = \sum_{z=0}^{\infty}\mathbb P(Z_i=z) x^z.
$
%
%
%
%
%
%
 Conditioning on $Q_i$ and exploiting independence of $Q_i$ and $D_i$ (which follows from the independence of a user's activation from users' and base stations' placements), we have $\mathbb E\left[ x^{Z_i} \right]= \sum_{q=1}^\infty \Lambda_q \mathbb E\left[ x^{q D_i}\right]$. Using the latter formula, polynomial representation of $D_i$, it can be derived (it can be shown that the effects of boundary placements vanish) that $\mathbb E\left[ x^{Z_i} \right]$ in the asymptotic regime becomes:
\begin{equation}
\label{eq-characteristic-function}
\Gamma(x) := \sum_{q=1}^\infty \Lambda_q   e^{ -\delta (1-x^q)}.
\end{equation}
This is the asymptotic \emph{node-oriented} users' degree distribution.
 We will also need the edge-oriented distribution $\gamma(x)=\Gamma^\prime(x)/\Gamma^\prime(1)$, e.g.,~\cite{RichardsonUrbanke}.
 A straightforward calculation shows that:
$
\gamma(x) := \sum_{q=1}^\infty \frac{q\,\Lambda_q}{\lambda} x^{q-1} e^{ -\delta (1-x^q)}.
$
 It can be shown (see Appendix; see also, e.g.~\cite{liva}) that the (edge-oriented) degree distribution $\chi(x)$ for arbitrary fixed check node~$(B_l,t)$ is asymptotically:
 %
 %
$
\chi(x) := e^{-G \delta \lambda(1-x)}.
$

\vspace{-0mm}
\section{Decoding algorithm}
\label{sec-decoding-algorithm}
We now present the decoding algorithm. It assumes that each base station~$B_l$ knows the IDs of all of its adjacent users~$U_i$ (which can be done through some sort of association procedure); for each $U_i$, $B_l$ also knows all the~$U_i$'s adjacent stations.
%
 The decoding algorithm is iterative and all base stations operate in the same way. Each base station $B_l$ maintains over iterations~$s$ a vector $z_l$ of $\tau$ signals, where the $t$-th entry $z_{l,t}$ serves as a current superposition signal of $B_l$ at slot $t$. Each iteration~$s$ has three steps, detailed below. We set the number of iterations to~$\tau m$.


{\emph{Step~1: Temporal SIC and Transmit}}. (\emph{Temporal SIC}) Station $B_l$ works with its local $\tau$ slots (check nodes) and performs standard SIC across them.  Namely, it checks if there are any singleton local slots. If there exists one, $B_l$ detects the signal from the corresponding user, $U^{(l)}$, collects this user and removes its contribution in all the remaining local slots where $U^{(l)}$ was active; we symbolically write this as $z_{l,t} \leftarrow z_{l,t} -  U^{(l)}$, for all $t$ where $U^{(l)}$ was active. The temporal SIC lasts until there are no more singleton local slots at $B_l$. (\emph{Transmit}) Station~$B_l$ forms the list $\mathcal U^{(l),\mathrm{out}}$ of all collected users during temporal SIC at the current iteration~$s$. For each $U^{(l)}$ in $\mathcal U^{(l),\mathrm{out}}$, $B_l$ broadcasts the information packet from $U^{(l)}$ (together with the $U^{(l)}$'s ID and its activation slots list) to all base stations adjacent to $U^{(l)}$. Go to step~2.

{\emph{Step~2: Check termination}}. If either all slots (check nodes) at $B_l$ are resolved or $s=\tau m$, $B_l$ leaves the algorithm. Otherwise, it goes to step~3.

{\emph{Step~3: Receive and Spatial ICs}}.  (\emph{Receive}) Station $B_l$ constructs the set $\mathcal U^{(l),\mathrm{in}}$  of all distinct users that it received at step~1. If $\mathcal U^{(l),\mathrm{new}}:=\mathcal U^{(l),\mathrm{in}} \setminus
\mathcal U^{(l),\mathrm{out}} = \emptyset$, set $s\leftarrow s+1$ and go to step~2. (\emph{Spatial ICs}) Otherwise, for each $U^{(k)}$ in $\mathcal U^{(l),\mathrm{new}}$, $B_l$ subtracts the contribution of $U^{(k)}$ at all its local slots (check nodes) where $U^{(k)}$ was active, $z_{l,t}\leftarrow z_{l,t} - U^{(k)}$.
 Set $s \leftarrow s+1$ and go to step~1.


Similarly to, e.g.,~\cite{liva},
the decoding algorithm can be represented as the evolution of graph~$\mathcal G$. At each iteration~$s$,
we remove from~$\mathcal G$ all the users $U_i$ collected at~$s$. (User~$U_i$ is collected at~$s$ if it is adjacent
to at least one check node that has degree equal to one at~$s$.) Also, for each collected
$U_i$, we remove all its incident edges, and all the check nodes $(B_l,t)$ that collected $U_i$ at~$s$.
 It can be shown that, with the described graph-cleaning process, no additional users are collected at
 iterations $s>\tau m$; this is why we set in our algorithm the maximal number of iterations be~$\tau m$.
See Figure~\ref{figure-example} (top right and bottom) for an example. 

\section{Performance analysis}
\label{section-performance-analysis}
Subsection~\ref{subsection-perf-metrics} sets up our analysis and introduces relevant performance metrics,
and Subsection~\ref{subsection-results} presents our results.
\subsection{Setup and performance metrics}
\label{subsection-perf-metrics}
We introduce four metrics relevant to our studies: expected fraction of collected users,
normalized throughput, maximal normalized load, and threshold normalized load.
 Denote by $1_{\mathcal S}$ the indicator function of event~$\mathcal S$.
 The expected fraction of collected users is given by:
$
\mathbb E \left[ \frac{1}{n}\sum_{i=1}^n 1_{\{U_i\,\mathrm{coll.}\}}\right] $,
and due to users' symmetry, it equals
$\mathbb P \left( U_i\,\mathrm{coll.} \right)$, for arbitrary fixed user~$U_i$.
 Next, we define the normalized throughput as the expected number of collected users per base station, per slot:
$
T(G)= \frac{n\,\mathbb P \left( U_i\,\mathrm{coll.} \right)}{\tau\,m}$ $ = G \mathbb P \left( U_i\,\mathrm{coll.} \right).
$
  Further, given a small positive number $\epsilon$, we define the maximal load $G^\bullet(\delta,\epsilon)$ by:
$
G^\bullet(\delta,\epsilon):=$ $\sup\left\{ G \geq 0:\, \mathbb P \left( U_i\,\mathrm{coll.} \right) \geq 1-\epsilon\right\}.
$
 In words, $G^\bullet(\delta,\epsilon)$ is the largest normalized load for which a fixed user
is collected with a guaranteed ``quality of service'', i.e., with probability at least $1-\epsilon.$
 Recall that $\mathbb P \left( U_i\,\mathrm{coll.} \right) \leq \mathbb P \left( U_i\,\mathrm{cov.} \right)$. Thus, when $\mathbb P \left( U_i\,\mathrm{cov.} \right)<1-\epsilon$, there is no value of load $G$ that
yields $\mathbb P \left( U_i\,\mathrm{coll.} \right)$ greater than or equal $1-\epsilon$.
  In such case, by convention, we set $G^\bullet(\delta,\epsilon)=0.$ Next, consider
  the asymptotic setting. We have $\mathbb P \left( U_i\,\mathrm{cov.} \right) \rightarrow 1-\mathrm{exp}(-\delta)$, and hence the maximal possible $\mathbb P \left( U_i\,\mathrm{coll.} \right)$ equals $1- \mathrm{exp}(-\delta)$.
   We define the threshold on the normalized load
 $G^\star(\delta)$ as the largest normalized load that allows for maximal possible $\mathbb P \left( U_i\,\mathrm{coll.} \right)$:
 \begin{equation}
\label{eqn-threshold}
G^\star(\delta):=\sup\left\{ G \geq 0:\, \lim_{n\rightarrow \infty}\mathbb P \left( U_i\,\mathrm{coll.} \right) = 1-e^{-\delta}\right\}.
\end{equation}
To benchmark our multiple base station system, we consider a single base station system with $\tau$ slots per frame,
with the base station placed at the center of the region $\mathcal A$.
 Base station's radius is sufficiently large such that it hears each of the $n$ users.
 Users perform the same transmission protocol as with the multiple base station system, i.e.,
 they select slots according to distribution $\Lambda_q$. Base station performs
 the interference cancellation decoding as in~\cite{liva}. There exists a threshold $H^\star$
 on the load $H=n/\tau$ below which $\mathbb P \left( U_i\,\mathrm{coll.} \right) \rightarrow 1$
  as $n \rightarrow \infty$, $\tau =\tau(n) \rightarrow \infty$,~\cite{liva}. The quantity $H^\star$ will be useful
  in our subsequent analysis.

\subsection{Results}
\label{subsection-results}
We first consider the threshold $G^\star(\delta)$ in~\eqref{eqn-threshold} with our multiple base station system.
 Theorem~\ref{theorem-threshold-lower-bound} shows existence of a non-trivial (strictly positive) $G^\star(\delta)$ and gives a lower bound on its value. Proof of Theorem~\ref{theorem-threshold-lower-bound} is in the Appendix.
%
%
%
\begin{theorem}
\label{theorem-threshold-lower-bound}
Consider the decoding algorithm in Section~\ref{sec-decoding-algorithm} in the asymptotic setting when
$n \rightarrow \infty$, $m = m(n)\rightarrow \infty$, $\tau=\tau(n) \rightarrow \infty$,
$r=r(n) \rightarrow 0,$ such that $m r^2 \pi \rightarrow \delta$, and $\frac{n}{\tau m} \rightarrow G$, for positive constants
$\delta$ and $G$. Let users select slots according to degree distribution $\Lambda_q$, and let
$H^\star$ be the corresponding threshold for the single base station system. Then, $G^\star(\delta)$ in~\eqref{eqn-threshold} satisfies:
$G^\star(\delta) \geq \frac{1}{8 \,e} \frac{H^\star}{\delta}$.
\end{theorem}
We now interpret Theorem~\ref{theorem-threshold-lower-bound} and use it
to compare our multiple base station system with the single base station system.
We first consider the peak normalized throughput $T^\star(\delta)=\max_{G \geq 0}T(G)$.
By definition of $G^\star(\delta)$, for all values of $G$ below $G^\star(\delta)$, $T(G)$
asymptotically equals $T(G)=G(1-\mathrm{exp}(-\delta))$. Hence,
by Theorem~\ref{theorem-threshold-lower-bound},
$T^\star(\delta) \geq
\frac{H^\star}{8\,e} \frac{1-\mathrm{exp}(-\delta)}{\delta}$.
 For the $1-\epsilon$ coverage, we need
 $\delta \geq \mathrm{ln}(1/\epsilon)$;
 the maximum of $\frac{H^\star}{8\,e} \frac{1-\mathrm{exp}(-\delta)}{\delta}$ over $\delta \geq \mathrm{ln}(1/\epsilon)$ is
 attained at $\mathrm{ln}(1/\epsilon)$
 and equals $c:=\frac{H^\star}{8\,e} \frac{1-\epsilon}{\mathrm{ln}(1/\epsilon)}$.
 Hence, the expected number of decoded users per slot, across \emph{all} base stations (unnormalized throughput),
 is at least~$c\times m$. Thus, compared with the single base station system, whose maximal
 throughput is a constant (roughly equal to $H^\star$), the $m$-base station system
 indeed achieves linear gains in~$m$.
 This conclusion extends our previous results in~\cite{MASSAP2}. Namely, we showed in~\cite{MASSAP2}
  that $m$-base stations also introduce linear gains over the single base station, when the corresponding two decoding algorithms do not employ
  temporal SIC. Here, we show that the same effect holds in the presence of temporal SIC as well.

Next, note that the lower bound on $G^\star(\delta)$ in Theorem~\ref{theorem-threshold-lower-bound}
is a decreasing function of~$\delta$. However, there is a tradeoff on the choice of $\delta$ (and hence, on the
communication radius~$r$) with respect to the system performance, measured, e.g., by the maximal load~$G^\bullet(\delta,\epsilon)$.
 We illustrate this tradeoff in Figure~\ref{figure-throughput} (top), where we plot the simulated value of $\mathbb P(U_i\,\mathrm{coll.})$ versus~$G$ for different values of
 $\delta \in \{3,5,12\}$, for the system with $m=40$ base stations
 and $\tau=40$ slots. (We refer to Section~\ref{section-numerical} for simulation details.) On one hand, we can see that the peak of $\mathbb P(U_i\,\mathrm{coll.})$ (equal to $1-\mathrm{exp}(-\delta)$)
  increases with the increase of $\delta$; on the other hand, the threshold $G^\star(\delta)$
   decreases, as predicted by Theorem~\ref{theorem-threshold-lower-bound}. (Intuitively,
   too large $\delta$ eliminates the benefits of spatial diversity.) For a given
   target ``quality of service'' (decoding probability) $1-\epsilon$, there is an optimal $\delta$
    that strikes the balance between the two above effects.

\textbf{And-or-tree heuristic for evaluation of $\mathbb P(U_i\,\mathrm{coll.})$}.
Recall the users' node-oriented degree distribution $\Gamma(x)$, users' edge-oriented degree distribution $\gamma(x)$,
and the check nodes' edge-oriented degree distribution $\chi(x)$ in the final paragraph of Section~\ref{section-model}.
 We apply standard and-or-tree iterations, e.g.,~\cite{liva}, with
 degree distributions $\Gamma(x), \gamma(x)$, and $\chi(x)$,
 and the maximal number of iterations $S$; in our simulations, we set $S=\tau\,m$.
 We estimate $\mathbb P (U_i\,\mathrm{coll}.)$ as $\mathbb P (U_i\,\mathrm{coll}.) \approx 1-\Gamma(p_S)$,
 where $p_S$ is the output of the and-or-tree evolution, initialized by $p_0=q_0=1$, and iterations:
\begin{equation}
q_s = \gamma(p_{s-1}),\:\:p_s = 1-\chi(1-q_s), \:s=1,...,S.
\end{equation}
\begin{remark}
\label{remark-and-or-tree} The spatio-temporal decoding graph here has a different structure from the classical random graph structure
in, e.g.,~\cite{liva}. For example, a fixed user $U_i$ here cannot connect to check nodes (base stations) outside its communication radius~$r$, and
hence many user-check node links are precluded. Because of this, there is no guarantee that the and-or-tree analysis gives
exact formulas for decoding probability in the asymptotic setting (in the sense that, for the maximal number of
decoding iterations set to~$S$, we have
$\lim_{n \rightarrow \infty} \mathbb P(U_i\,\mathrm{coll.})=1-\Gamma(p_S)$.) Nonetheless, we
apply the and-or-tree analysis as if the graph were a standard random graph. Simulations in Section~\ref{section-numerical}
show that the and-or-tree heuristic follows well the actual system performance.
\end{remark}
\vspace{-0mm}
\section{Numerical studies}
\label{section-numerical}

In this section, we study the performance of our spatio-temporal cooperative decoding through simulations.
Specifically, we assess the benefits of introducing temporal SIC, spatial SIC,
and simultaneous spatio-temporal SIC (as proposed here). That is, we compare four
different decoding algorithms, each of which we briefly outline. (1)~Non-cooperative decoding: each base station performs standard slotted Aloha decoding in isolation:
base station $B_l$ decodes a user at slot $t$ if and only if it receives a clean signal at slot~$t$. (2)~Spatial SIC: decoding is completely decoupled across slots. At each slot~$t$,
base stations perform spatial SIC, similarly to step~3 of the decoding algorithm proposed here (see also~\cite{MASSAP2}).
(3)~Temporal SIC: each base station works in isolation from other stations and
performs (temporal) SIC as in~\cite{liva}. (4)~Spatio-temporal SIC: this is the proposed algorithm in Section~\ref{sec-decoding-algorithm}. To explore the benefits of spatial and temporal SICs on a common ground, we use with each decoding the same users' transmission protocol, i.e., the same~$\Lambda_q$. Clearly, distribution~$\Lambda_q$ (in principle) could be optimized for each scenario, and the optimal distribution is likely to be different for each scenario. Another meaningful comparison of decoding algorithms would be in terms of their optimized distributions; this is not considered here.
\begin{figure}[thpb]
      \centering
       \includegraphics[height=2 in,width=3.5 in]{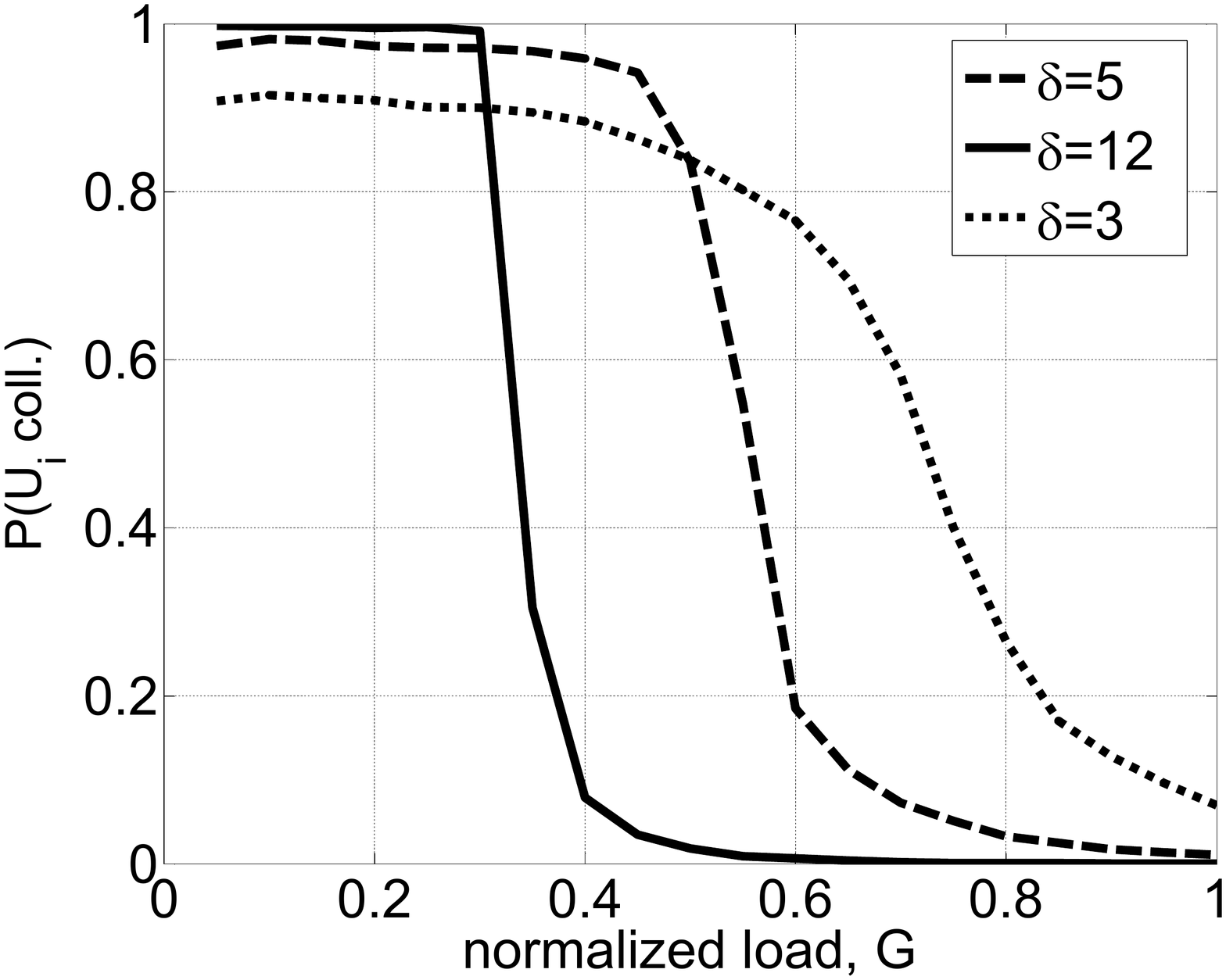}
       \includegraphics[height=2 in,width=3.5 in]{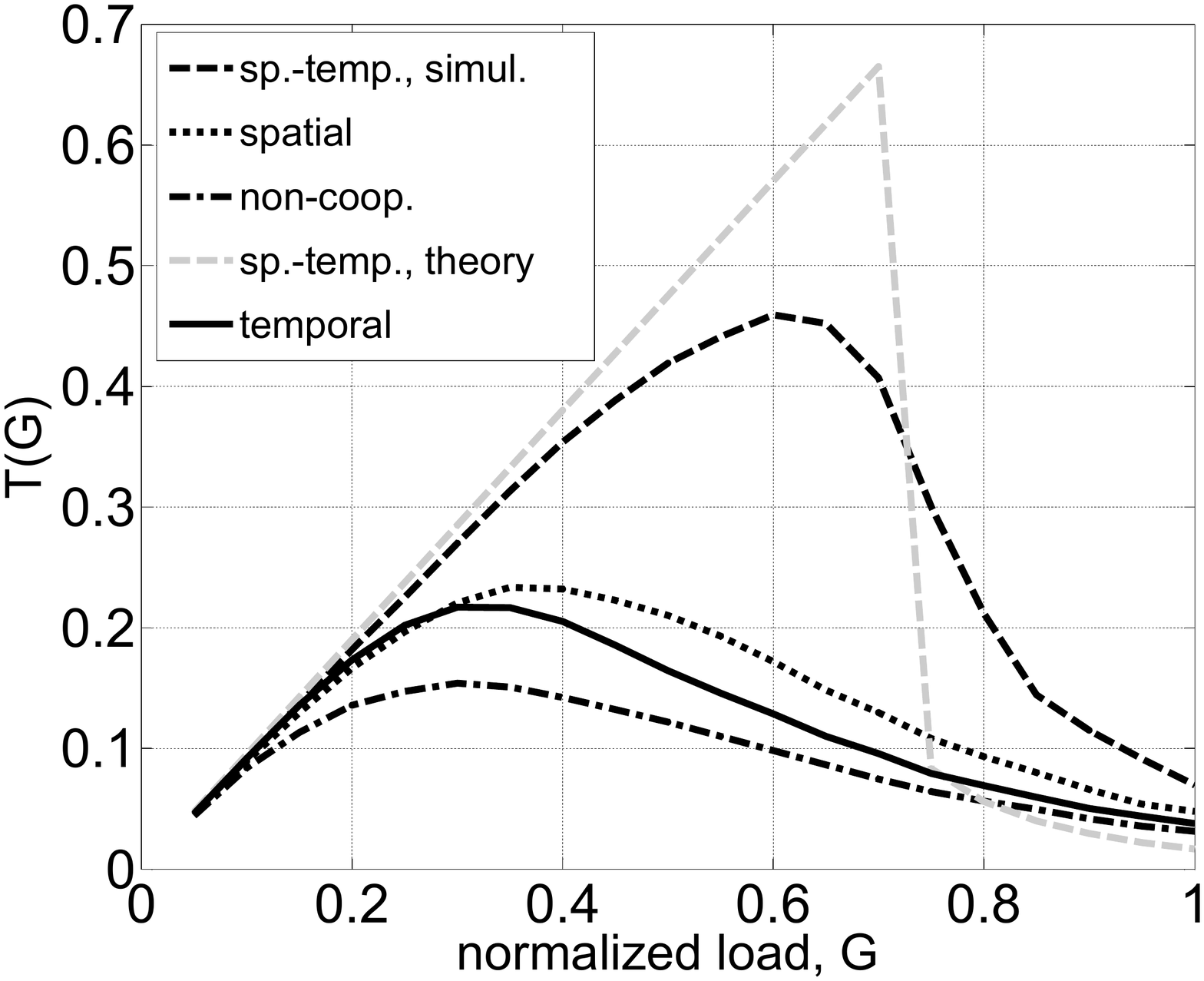}
       \includegraphics[height=2 in,width=3.5 in]{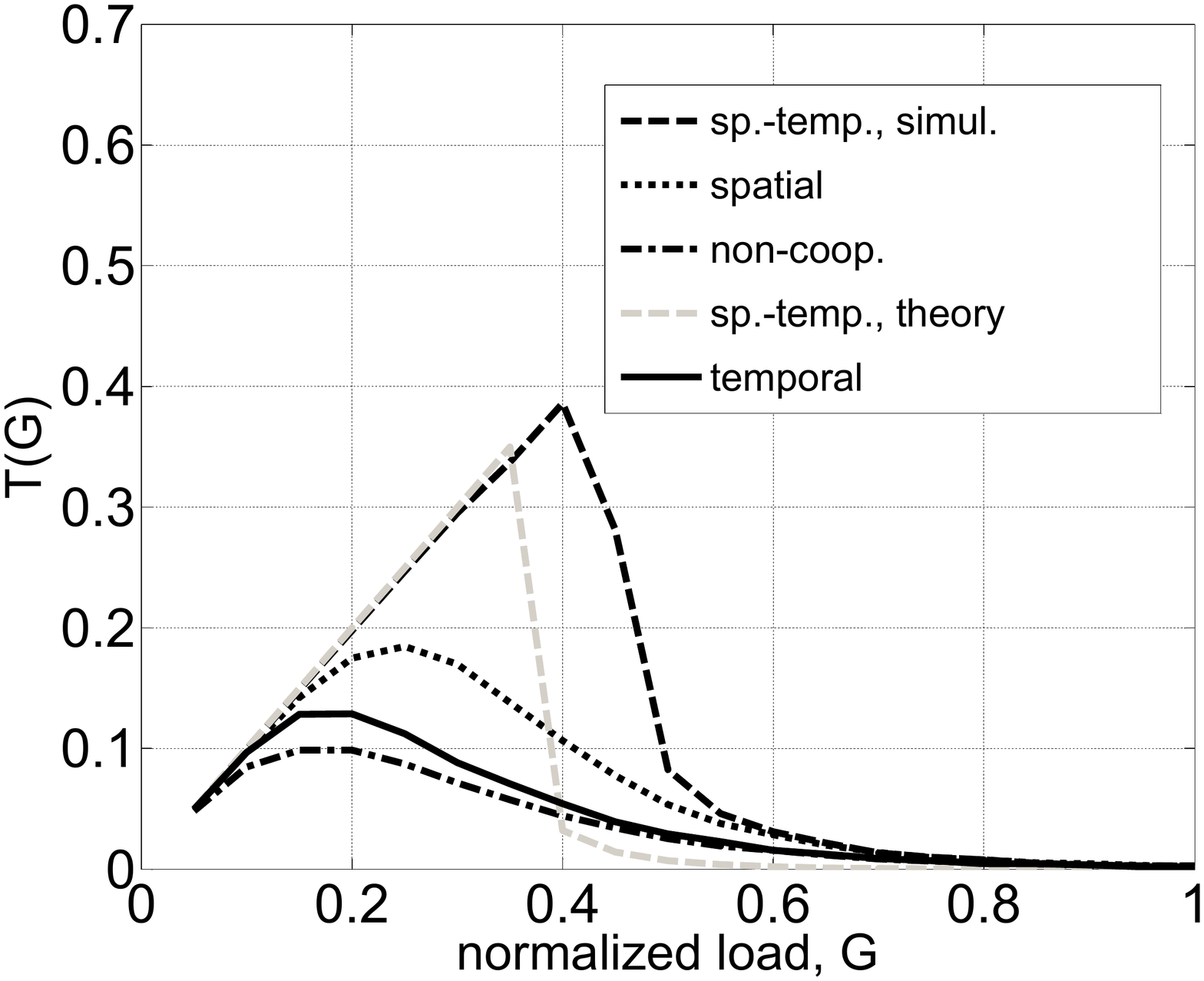}
       \caption{Top Figure: Simulated $\mathbb P(U_i\,\mathrm{coll.})$ versus $G=n/(\tau m)$
       with spatio-temporal decoding for different values of $\delta=3,5,12$ for the system
       with $m=40,\tau=40$. Two bottom plots: Simulated normalized
       throughout $T(G)$ versus $G$ for four decoding algorithms;
       second Figure from top: $\delta=3$; third Figure: $\delta=7$.}
       \label{figure-throughput}
       \vspace{-0mm}
\end{figure}
%
%
%
%
%
%
%
Throughout simulations, we use the following setup. We consider the system with $m=40$ base stations
and $\tau=40$ slots. We simulate the normalized throughput $T(G)$ for different values of $G=n/(\tau m)$ by varying $n$
 such that $G$ varies within~$[0,1]$. We evaluate
  $T(G)$ with the four decoding algorithms via Monte Carlo simulations. For each fixed~$n$ (each fixed~$G$),
 we perform $30$ simulation runs, i.e.,
 we generate $30$ different graphs (with $30$ different users' and stations' placements) and perform four decoding algorithms over
 each graph. With all four algorithms, we estimate $T(G)$ as the total number of distinct
 collected users during $\tau$ slots (across all base stations and all $\tau$ slots) divided by~$m\tau$.
     With all four algorithms, users transmit across slots
    according to degree distribution $\Lambda_2=1$, $\Lambda_q=0$, $q \neq 2$. This is a representative
    degree distribution from the literature, proposed in~\cite{SlottedALOHAwithIC} (and it is not optimized).

Figure~\ref{figure-throughput} plots the normalized throughput for four algorithms
and different values of $\delta$ (second from top Figure: $\delta=3$, third from top: $\delta=7$.)
 Both values for $\delta$ ensure at least $95\%$ coverage probability.
 First, we can see that, for both $\delta$'s, spatio-temporal SIC shows significant gains over
 the remaining three methods. For example,
 for $\delta=3$, spatio-temporal SIC achieves peak throughput about~$0.45$, spatial SIC--
 about $0.23$, temporal SIC--about $0.22$, and non-cooperative--about $0.16$. Thus, performing jointly spatial and temporal SIC gives much better peak throughput than performing solely either temporal or spatial SIC.
 It is also interesting to compare our spatio-temporal decoding with the
 single base station system that also uses $\Lambda_2=1$ and decoding in~\cite{liva},
 in terms of the \emph{unnormalized} throughput, cumulative over $m=40$ base stations.
 With $\delta=3$, our decoding achieves about $0.45 \times m$, while the single base station system
 achieves about~$0.55.$ Hence, introducing $m=40$ base stations increases the unnormalized throughput
 about $32$ times with respect to the single base station system. Next, we can see that both peak throughput and threshold are smaller for
  the larger value of~$\delta$, as predicted by Theorem~\ref{theorem-threshold-lower-bound}. Finally, we can see that our heuristic for throughput based on the and-or-tree analysis follows well the trend of the simulated system performance.


\vspace{-0mm}

\section{Conclusion}
\label{section-conclusion}
We considered framed slotted Aloha where $m$ base stations cooperate
to decode the messages from $n$ users. Users and base stations are randomly deployed over a unit area.
 Users transmit their packet replicas at multiple slots (out of $\tau$ available slots in each frame), selected according to a
 prescribed distribution~$\Lambda_q$. Users' messages are simultaneously heard by all base stations within distance~$r$.
 We propose a cooperative decoding that utilizes both spatial and temporal (successive) interference cancellation
 mechanisms~(SICs). The spatial SIC allows for cleaning of interference across base stations, through communication among neighboring base stations. The temporal SIC allows for cleaning of interference across different slots, locally at each base station.
 We established that there exists a threshold on the normalized load $G=n/(\tau m)$ below which
 the decoding probability converges asymptotically to the maximal possible value, equal to the probability that a
 user is head by at least one station; and we found a lower bound on this threshold. Further, we gave a heuristic
 for the evaluation of decoding probability based on the and-or-tree analysis. Our results
 reveal that $m$ base stations introduce $m$-fold (linear) gains in the throughout,
 when compared with a single base station system where users transmit according to the same distribution~$\Lambda_q$.


\vspace{-0mm}

\bibliographystyle{IEEEtran}
\bibliography{IEEEabrv,bibliography}

\appendix

\subsection{Degree distributions in graph~$\mathcal G$}
\label{subsection-appendix-degree-dist}

\textbf{Auxiliary arguments for derivation of $\Gamma(x)$ in~\eqref{eq-characteristic-function}}.
Recall from Section~\ref{section-model} that the polynomial representation of
the users' degree $Z_i=Q_i\,D_i$ is:
\begin{equation}
\label{eqn-app-poly-1}
\mathbb E [x^{Z_i}]=\sum_{q=1}^\infty \Lambda_q \mathbb E\left[ x^{q D_i}\right],
\end{equation}
for finite $m=m(n),n,r=r(n), \tau=\tau(n)$.
We want to find $\lim_{n \rightarrow \infty}\mathbb E [x^{Z_i}]$, for all $x \in \mathbb R$,
when $m(n) \rightarrow \infty$, $r(n) \rightarrow 0$, and
$m r^2 \pi \rightarrow \delta$, $\delta>0.$
 We will show that:
 \begin{equation}
\label{eqn-app-poly-8}
\mathbb E[x^{q D_i}] \rightarrow e^{-\delta(1-x^q)},\,\,\forall x\in \mathbb R,\,\,\forall q\in\{1,2,...\}.
 \end{equation}
Fix arbitrary $x \in \mathbb R$, and fix arbitrary $q\in \{1,2,...\}$.
We evaluate $\mathbb E\left[ x^{q D_i}\right]$ by
conditioning on either nominal or boundary placements:
\begin{eqnarray}
\label{eqn-app-poly-2}
\mathbb E[x^{q D_i}] &=& \mathbb E[x^{q D_i}\,|\,u_i\in \mathcal{A}^{\mathrm{o},r}]\,
\mathbb P (u_i \in \mathcal{A}^{\mathrm{o},r})\\
&+&
\mathbb E[x^{q D_i}\,|\,u_i\in \partial \mathcal{A}^{r}]\,
\mathbb P (u_i \in \partial \mathcal{A}^{r}). \nonumber
\end{eqnarray}
Conditioned on the nominal placement, $D_i$ is a binomial random variable
with the number of trials $m$ and success probability $r^2\pi$,
$\mathrm{Binomial}(m,r^2\pi)$ for short.
Calculating the polynomial representation for $\mathrm{Binomial}(m,r^2\pi)$:
\begin{equation}
\label{eqn-app-poly-3}
\mathbb E[x^{q D_i}\,|\,u_i\in \mathcal{A}^{\mathrm{o},r}]  = (1+(x^q-1)r^2\pi)^m.
\end{equation}
Now, consider a boundary placement $u_i=\kappa$, $\kappa \in \partial \mathcal{A}^{r}$.
Conditioned on $u_i=\kappa$, $D_i$ is $\mathrm{Binomial}(m,\pi(\kappa))$,
where the success probability $\pi(\kappa)$ satisfies:
$r^2\pi/4 \leq \pi(\kappa) \leq r^2\pi$, $\forall \kappa \in \partial \mathcal{A}^{r}$. Thus, for any
$\kappa \in \partial \mathcal A^r$:
\begin{equation}
\label{eqn-app-poly-3}
\mathbb E[x^{q D_i}\,|\,u_i = \kappa ]  = (1+(x^q-1)\pi(\kappa))^m.
\end{equation}
Next, note that $|1+(x^q-1)\pi(\kappa)| \leq 1 + |x^q-1| \pi(\kappa)
\leq 1+ |x^q-1|r^2\pi$, $\forall \kappa \in \partial \mathcal A^r$.
Using, this, we get:
 $|(1+(x^q-1)\pi(\kappa))^m|  = |1+(x^q-1)\pi(\kappa)|^m
  \leq (1+(x^q-1)r^2\pi)^m$, $\forall \kappa \in \partial \mathcal A^r$.
  The last inequality means that:
\begin{eqnarray*}
- (1+(x^q-1)r^2\pi)^m &\leq& \mathbb E[x^{q D_i}\,|\,u_i = \kappa ]  \\
&\leq& (1+(x^q-1)r^2\pi)^m,\,\,\forall \kappa \in \partial \mathcal A^r,
\end{eqnarray*}
and therefore, integrating over $\kappa \in \partial \mathcal A^r$:
\begin{eqnarray}
\label{eqn-app-poly-5}
- (1+(x^q-1)r^2\pi)^m &\leq& \mathbb E[x^{q D_i}\,|\,u_i\in \partial \mathcal A^r ]  \\
&\leq& (1+(x^q-1)r^2\pi)^m. \nonumber
\end{eqnarray}
Now, pass to the asymptotic setting $m \rightarrow \infty$, $r \rightarrow 0$, $mr^2\pi \rightarrow \delta$, $\delta>0$.
 We have that:
 \begin{equation}
 \label{eqn-app-poly-6}
 (1+(x^q-1)r^2\pi)^m \rightarrow e^{-\delta(1-x^q)}.
 \end{equation}
Now, using
 $\mathbb P(u_i\in \partial \mathcal A^{r}) = 1 -(1-4r)^2 \rightarrow 0$,
 \eqref{eqn-app-poly-5} and \eqref{eqn-app-poly-6}, we obtain that:
\begin{equation}
\label{eqn-app-poly-7}
\mathbb E[x^{q D_i}\,|\,u_i\in \partial \mathcal{A}^{r}]\,
\mathbb P (u_i \in \partial \mathcal{A}^{r}) \rightarrow 0.
 \end{equation}
Finally, from \eqref{eqn-app-poly-2}, \eqref{eqn-app-poly-3}, \eqref{eqn-app-poly-6}, \eqref{eqn-app-poly-7},
and the fact that $\mathbb P(u_i\in  \mathcal A^{\mathrm{o},r}) = (1-4r)^2 \rightarrow 1$,
we obtain~\eqref{eqn-app-poly-8}, as desired. Substituting  \eqref{eqn-app-poly-8} in \eqref{eqn-app-poly-1},
we finally obtain~\eqref{eq-characteristic-function}.

\textbf{Derivation of check nodes' degree distributions in~$\mathcal G$}. Consider graph~$\mathcal G$ defined
in Section~\ref{section-model}, and fix an arbitrary check node~$(B_l,t)$.
Assume a nominal placement $b_l \in \mathcal A^{\mathrm{o},r}$.
As users' placements are independent, and they choose slots independently from each other and independently from the placements,
the degree of $(B_l,t)$ is a binomial random variable, with $n$ trials, and success probability
 equal $\mathbb P(U_i \mathrm{\,adjacent\,to\,}B_l\,|\,b_l \in \mathcal A^{\mathrm{o},r}) \,\mathbb P(U_i \mathrm{\,active\,at\,}t\,|\,b_l \in \mathcal A^{\mathrm{o},r})$.
 We have
 $\mathbb P(U_i \mathrm{\,adjacent\,to\,}B_l\,|\,b_l \in \mathcal A^{\mathrm{o},r})=r^2\pi$.
 We now evaluate $\mathbb P(U_i \mathrm{\,active\,at\,}t\,|\,b_l \in \mathcal A^{\mathrm{o},r})$ by conditioning on the number of activation slots $Q_i$ of $U_i$. The latter probability equals: $\sum_{q=1}^\infty \frac{q}{\tau}\Lambda_q = \lambda/\tau$.
 Hence, conditioned on the nominal placement, the success probability equals $r^2 \pi \lambda/\tau$ and the mean of
 the check node's degree is $\frac{n r^2 \pi \lambda}{\tau} = \frac{n r^2 \pi \lambda m}{\tau m} $,
 which converges to $G \delta \lambda$. Therefore, conditioned on the nominal placement,
 a check node's degree distribution converges to Poisson with mean $G \delta \lambda$, i.e.,
 $\mathbb P(\mathrm{degree}(B_l,t)=d\,|\,b_l\in \mathcal A^{\mathrm{o},r}) \rightarrow
 e^{-G \delta \lambda}(G \delta \lambda)^d \d!$, $d=0,1,...$
 Using the latter, the total probability formula
 $\mathbb P(\mathrm{degree}(B_l,t)=d) = $
 $\mathbb P(\mathrm{degree}(B_l,t)=d\,|\,b_l\in \mathcal A^{\mathrm{o},r})\mathbb P(b_l\in\mathcal A^{\mathrm{o},r}))$
 $+$
 $\mathbb P(\mathrm{degree}(B_l,t)=d\,|\,b_l\in \partial \mathcal A^{r})\mathbb P(b_l\in\partial \mathcal A^{r}))$,
 and the fact that $\mathbb P(b_l\in\mathcal A^{\mathrm{o},r})) \rightarrow 1$,
 $\mathbb P(b_l\in\partial \mathcal A^{r})) \rightarrow 0$, we finally obtain:
  $\mathbb P(\mathrm{degree}(B_l,t)=d) \rightarrow
 e^{-G \delta \lambda}(G \delta \lambda)^d \d!$, $d=0,1,...$ Therefore, the node-oriented distribution of check nodes is asymptotically
 Poisson with mean $G \delta \lambda$. It is easy to show that the edge-oriented distribution is then the same, e.g.,~\cite{liva},
 and the corresponding polynomial representation $\chi(x)=\mathrm{exp}(-G\delta\lambda(1-x)).$

\subsection{Proof of Theorem~\ref{theorem-threshold-lower-bound}}
\label{subsection-appendix-proof}
Fix an arbitrary user $U_i$. To prove the Theorem, we will
devise a useful lower bound on $\mathbb P(U_i\,\mathrm{coll.})$, in the asymptotic setting.
 Consider the placement $u_i=q$. Recall the nominal placement set
$\mathcal{A}^{o,r}$. Because $r\rightarrow 0$ as $n \rightarrow \infty$, it suffices to lower bound
$\mathbb P(U_i\,\mathrm{coll.}\,|\,u_i=q)$ for any $q \in \mathcal{A}^{o,r}$. (We strictly show why this is sufficient later in the proof.)
Let $\mathbf{B}(\zeta,\rho)$ denote the disc centered at $\zeta$ with radius $\rho$. Further, denote by $N_B(u_i)$ the number of base stations in $\mathbf{B}(u_i,r)$,
and by $N_U(u_i)$ the number of users different than $U_i$ in $\mathbf{B}(u_i,2r)$.
 We first explain the intuition behind the proof, and then we formalize it through equations.
 We construct a specific scenario when $U_i$ is collected and evaluate its probability.
 The scenario is as follows: user $U_i$ has at least one base station in its $r$-neighborhood ($N_B(u_i) \geq 1$),
 and there are at most $C$ users different than $U_i$ in the $U_i$'s $2r$-neighborhood ($N_U(u_i) \leq C$).
 Without loss of generality, let $B_1$ be one of the base stations in $\mathbf{B}(u_i,r)$. In the considered scenario, $B_1$
  has in its neighborhood at most $C+1$ users. Then, the probability that $U_i$ is collected is greater than or equal
  the probability that $U_i$ is collected by $B_1$ that works as a single base station system (in the sense of the system described in
   the final paragraph of Subsection~\ref{subsection-perf-metrics})
   with $C+1$ users, i.e., with load $H=(C+1)/\tau$.

We now proceed with formalizing the above idea. We have:
{\small{
\begin{eqnarray*}
&\,&\mathbb P(U_i\,\mathrm{coll.}\,|\,u_i=q) \\
&\geq& \mathbb P(U_i\,\mathrm{coll.}\,|\,N_B(u_i) \geq 1,\,N_U(u_i)\leq C,\,u_i=q)\\
&\times& \mathbb P \left(  N_B(u_i) \geq 1,\,N_U(u_i) \leq C\,|\,u_i=q       \right).
\end{eqnarray*}}}
Next, note that:
\begin{eqnarray*}
&\,& \mathbb P \left(  N_B(u_i) \geq 1,\,N_U(u_i) \leq C\,|\,u_i=q       \right)\\
&=& \mathbb P \left(  N_B(q) \geq 1,\,N_U(q) \leq C\,|\,u_i=q       \right) \\
&=&\mathbb P \left(  N_B(q) \geq 1\right) \mathbb P \left(  N_U(q) \leq C\right),
\end{eqnarray*}
where the last equality holds by the independence of the users' and base stations' placements.
 Denote by $ \mathbb P(U_i\,\mathrm{coll.}\,|\,N_B(u_i) \geq 1,\,N_U(u_i)\leq C,\,u_i=q) = \widehat{P}$.
  We have:
\begin{eqnarray}
\label{eqn-proof-1}
\mathbb P(U_i\,\mathrm{coll.}\,|\,u_i=q)
\geq \widehat{P}\,\mathbb P \left(  N_B(q) \geq 1\right) \,\mathbb P \left(  N_U(q) \leq C\right).
\end{eqnarray}
Note that $N_B(q)$ is a binomial random variable with the number of trials
equal $m$ and success probability $r^2\pi$. Similarly, $N_U(q)$ is a binomial random variable with the number of trials
equal $n-1$ and success probability $4 r^2\pi$, and $\mathbb E[N_U(q)]=4 r^2\pi (n-1)$. From now on, we set
$C=\theta 4 r^2 \pi(n-1)$, for some $\theta >0$ that we specify later. We proceed by separately lower bounding each of the three probabilities on the right hand
side of~\eqref{eqn-proof-1}.

\textbf{Lower bound on $\widehat{P}$}. As explained in the intuition above, we have $\widehat P \geq P_{\mathrm{single}}$,
 where $P_{\mathrm{single}}$ is the probability that a fixed user $U_i$ is collected by the single base station system
 with $C+1=\theta(n-1)4 r^2\pi+1$ users, users' degree distribution $\Lambda_q$, and load
 $H=(C+1)/\tau$. (The term $1$ in $C+1$ comes from the inclusion of $U_i$ as well.) Note that we use here
 the fact that decoding probability with the single base station system
 is a monotonically non-increasing function of load $H$. (Conditioned on
 the number of served users be at most $C+1$, the worst case occurs for the number of users equal~$C+1$.)
 We know that, when $H \leq H^\star$, $P_{\mathrm{single}} \rightarrow 1$ as $n \rightarrow \infty.$
  We pick the largest $\theta>0$ such that the latter is fulfilled, i.e., such that
  $\frac{C+1}{\tau}=\frac{\theta(n-1)4r^2\pi }{\tau} + \frac{1}{\tau}=H^\star$.
  Using $G=n/(\tau m)$ and $\delta=m r^2 \pi$, it is straightforward to show that this translates into $\theta =  \frac{1}{4}\frac{H^\star}{\delta G}+o(1)$. Hence, we conclude that:
  \begin{equation}
  \label{eqn-proof-combine-1}
  \mathrm{for}\,\,\theta=\frac{1}{4}\frac{H^\star}{\delta G}:\,\,\, \lim_{n \rightarrow \infty} \widehat{P} = 1 .
  \end{equation}

\textbf{Lower bound on $\mathbb P(N_B(q) \geq 1)$}. Clearly, $\mathbb P(N_B(q) \geq 1)=\mathbb P(U_i\,\mathrm{cov.}\,|\,u_i=q)$, and hence:
 \begin{equation}
  \label{eqn-proof-combine-2}
  \lim_{n \rightarrow \infty} \mathbb P(N_B(q) \geq 1) = 1-e^{-\delta}.
  \end{equation}

\textbf{Lower bound on $\mathbb P(N_U(q) \leq \theta 4(n-1)r^2 \pi)$}.
We use the following inequality for a binomial random variable $Z$ with mean $\mu$, e.g.,~\cite{RandomizedAlgorithms}, Theorem~4.1:
 \[
 \mathbb P(Z> \theta\,\mu)< 2^{-\theta\,\mu}, \:\:\forall \theta \geq 2e.
 \]
Applying this inequality, we obtain:
\begin{equation}
\mathbb P(N_U(q) \leq \theta 4(n-1)r^2 \pi) \geq 1 - 2^{-\theta\,4(n-1)r^2\pi},\:\theta \geq 2e.
\end{equation}
Note that $(n-1)r^2\pi = n r^2 \pi - r^2 \pi
= n r^2 \pi \frac{\tau m}{\tau m} - r^2 \pi = G \delta \tau - r^2\pi \rightarrow +\infty
$, as $n\rightarrow \infty$ (because $\tau(n) \rightarrow \infty.$) Thus, we conclude that:
\begin{equation}
\label{eqn-combine-3}
\lim_{n \rightarrow \infty} \mathbb P(N_U(q) \leq \theta 4(n-1)r^2 \pi) = 1, \,\,\mathrm{for\,all}\,\,
\theta \geq 2 e.
\end{equation}
Now, we combine~\eqref{eqn-proof-combine-1}, \eqref{eqn-proof-combine-2}, and~\eqref{eqn-combine-3}.
Using~\eqref{eqn-proof-1}, we can see that, if $\frac{1}{4}\frac{H^\star}{\delta G} \geq 2 e$, i.e.,
if $G \leq \frac{1}{8\,e}\frac{H^\star}{\delta}$,
the following holds:
 $\liminf_{n \rightarrow \infty}
 \mathbb P(U_i\,\mathrm{coll.}\,|\,u_i=q) = 1-e^{-\delta},
 $
 for all $q \in \mathcal{A}^{o,r}$.
 Finally, note that $\mathbb P(u_i \in \mathcal{A}^{o,r} ) = (1- 4 r)^2$,
 which converges to one. Also,
 $\mathbb P(U_i\,\mathrm{coll.}) \geq \mathbb P(U_i\,\mathrm{coll.}\,|\,u_i  \in \mathcal{A}^{o,r})
 \mathbb P(u_i \in \mathcal{A}^{o,r} )$. Combining the last two observations,
 we obtain that
   $\liminf_{n \rightarrow \infty}
 \mathbb P(U_i\,\mathrm{coll.}) \geq  1-e^{-\delta},
 $
 if $G < \frac{1}{8\,e}\frac{H^\star}{\delta}$. Clearly,
 $\limsup_{n \rightarrow \infty} \mathbb P(U_i\,\mathrm{coll.}) \leq 1-e^{-\delta}$.
 Thus, we conclude that, if $G < \frac{1}{8\,e}\frac{H^\star}{\delta}$,
 then $\lim_{n \rightarrow \infty} \mathbb P(U_i\,\mathrm{coll.}) = 1-e^{-\delta}$, which
 means that the threshold $G^\star(\delta) \geq \frac{1}{8\,e}\frac{H^\star}{\delta}$. The proof is complete.
\end{document}